\documentclass[preprint,amsmath,amssymb,aps,prl,floatfix]{revtex4-2}
\usepackage{graphicx}
\usepackage{dcolumn}
\usepackage{bm}
\usepackage{float}
\usepackage{hyperref}

\begin{document}
\title{Establishing limits for the monopole production in pp accelerators}

\author{B. M. Carlos}
 \email{bruna.carlos@ufrgs.br}
\author{M. B. Gay Ducati}
 \email{beatriz.gay@ufrgs.br}
\affiliation{%
 High Energy Physics Phenomenology Group, Physics Institute,  Federal University of Rio Grande do Sul, \\
Caixa Postal 15051, CEP 91501-970, Porto Alegre, RS, Brazil
}%

\date{May 2021} 

\begin{abstract}
The spin $1/2$ magnetic monopole pair production and the spin $0$ monopolium production are studied in proton-proton collisions. The velocity-dependent coupling model and an additional effective coupling with a magnetic moment parameter are used to calculate the photon fusion and Drell Yan pair production cross sections. The monopole mass is unknown, hence the mass range employed here is based on the last results of ATLAS and MoEDAL experiments, which set a minimal value around 2 TeV. The cross sections are calculated for the LHC center of mass energies, its successor, the HE-LHC, and the future collider FCC. The estimated mass limits for observation in the LHC and the other accelerators are obtained and the advantages in using both effective couplings and the monopolium as an indirect measure are discussed.
\end{abstract}


\maketitle

\section{Introduction} \label{sec:outline}
One of the first formulations for a magnetic monopole was proposed by Dirac in the years 1931\cite{dirac1} and 1948\cite{dirac2}, and consisted of a point-like particle with non-specified mass and spin, carrying a magnetic charge $g$ and a non-physical string that carries the magnetic flux of the particle. This particle would be responsible for bringing symmetry to Maxwell's equations and explaining the charge quantization through the Dirac quantization condition (in SI natural units) 
\begin{equation}
    eg = 4\pi\frac{n}{2},
\end{equation}
where $e$ is the electron charge and $n$ a positive integer. New models for this unknown particle were proposed later on, like the dyon, a particle with both electric and magnetic charges theorized by Schwinger \cite{dyon}, and those in electroweak \cite{electroweak} and unification theories \cite{thooft}\cite{polyakov}\cite{GUT1}\cite{GUT2}\cite{GUT3}, but even with the support of several experiments\cite{exp1}\cite{exp2} none of them were discovered yet. A possible explanation for this lack of experimental evidence was first given in \cite{hill}, where it was assumed that due to the strong magnetic coupling the magnetic poles would always be in a bound state, called monopolium. These particles could have a much lower mass than the monopole alone and be produced with higher rates in accelerators.

The large value of the magnetic charge $g$ has lead to the development of effective models that alleviate the perturbative limits in order to make approximated predictions.
The simplest one assumes a velocity dependent coupling, where the moving monopole is treated as an electric charge and couples to the photon just like the electron. This idea was used to determine the first limits on Dirac's monopole mass\cite{betacoupling}\cite{exp1}, and has been followed since then in some theoretical works \cite{dougall},\cite{garcia3} and in the current experimental search for monopoles in pp colliders \cite{moedal}\cite{moedal2019}\cite{atlas2020}.
A more recent work \cite{baines} proposed the addition of a magnetic moment term to the usual velocity-dependent coupling, and this new parameter increases the limits where perturbative methods can be used.

Recent experiments dedicated to the monopole search \cite{atlas2020},\cite{moedal} have given lower bounds on the monopole mass for different spin and charges, considering the monopole production by photon fusion and Drell Yan in pp collisions. In this work, the monopole pair production is studied considering the two processes analyzed in the current experiments, assuming a Dirac's monopole with spin $1/2$ and employing the two effective coupling models cited above. The monopolium production is investigated for the photon fusion process. The mass bounds from recent results are considered for the estimates, with the objective to establish upper bounds for production in the LHC and the future colliders HE-LHC\cite{helhc} and FCC\cite{fcc}. 

In the next section the effective couplings are presented, followed by the formalism used in the processes of monopole and monopolium production, and the one for total cross sections in proton-proton collisions.

\section{The Monopole Production}\label{Monopole Production}
\subsection{Pair production}

In Dirac's formalism the coupling force between a monopole and an antimonopole is about $10^3$ times stronger than the one between an electron and a positron, which turns the usual perturbative methods of QED not directly applicable to monopoles. Facing the scenario where no formal theory to treat magnetic monopole interactions is fully developed, some effective couplings have been proposed, like the minimalistic model where the coupling is given by
$$
    \alpha_m = \frac{\beta^2g^2}{4\pi},
$$
with $\beta$ the monopole velocity. In this formulation the moving magnetic monopole is treated as an electric charge in analogy with the behaviour of a moving electron producing a magnetic field. With this coupling the monopole can be considered in the place of an electron, or other charged lepton, in many processes by the replacement $e \to g\beta$. The velocity dependence decreases the production rates, but now the coupling $\alpha_m$ can be perturbatively expanded in the limit $ \beta\ll 1$, and more predictions can be obtained.

An alternative to the velocity-dependent coupling is to add a consistent magnetic moment term dependence\cite{baines}. As the monopole itself generates a magnetic field, it will not be necessary an electromagnetic scattering to generate a magnetic moment, like in the electron case. Also, due to the large value of $g$, the monopole can be expected to have a great magnetic moment that would be relevant already in a tree-level diagram. A parameter $\kappa$ can be defined in terms of the magnetic moment of the monopole
\begin{equation}
    \mathbf{\mu_m}=\frac{g\beta}{2m}2(1+2\hat{\kappa})\mathbf{S},
\end{equation}
where $\hat{\kappa}=\kappa m$ and $\mathbf{S}$ is the particle spin. Now, the photon-monopole coupling will be proportional to $g\beta$ plus a term with $\kappa$ dependence, and the perturbative expansions are valid in the limit $\kappa\gg 1$, $\hat{\kappa}\gg 1$ and $\beta\ll 1$\cite{baines}.

The results obtained with these two couplings are only indicative, for they are based mostly in effective calculations. However, they can be used to make estimations about whether the monopoles are detectable in the energies of current accelerators or not.

\section{The Monopole Production}\label{Monopole Production}
\subsection{Pair production}\label{subsec Pair Production}

Although in Dirac's theory the monopole does not have a defined spin, here the case $\hat{S}=1/2$ is going to be considered in order to build a symmetry with the electron (for spin $0$ and $1$ monopoles, we refer to \cite{spins} and \cite{baines}). The monopole-antimonopole pair production can be studied in two processes: photon fusion (u and t channels) and Drell Yan, both in leading order to avoid the issues that come up with the expansion of the magnetic coupling. 

Due to the ultrarelativistic energies reached at LHC, the flux of photons emitted by protons plays an important role in the production of many particles, such as leptons or charged scalar particles that have been widely studied \cite{twophoton} and used to investigate the Higgs boson production\cite{higgs}\cite{gay} and particles beyond the Standard Model\cite{bsm}. In collisions involving protons, the Drell Yan mechanism\cite{dy} can also be used to investigate such particles, as it has been indicated\cite{drees} that for lepton and Higgs production those cross sections are $\sim 10^2$ times larger than those for photon fusion. However, when considering the magnetic coupling, the photon fusion cross section has an enhancement of $\sim g^2\beta^2$ as compared to Drell Yan, which makes this process a more relevant candidate to consider for the monopole production in virtue of the quantization condition.  

The production cross sections can be obtained by replacing the electron charge and the electromagnetic coupling by $g\beta$ and $\alpha_m$, respectively, in the expressions for electron-positron production for photon fusion\cite{drees}
\begin{equation}
    \hat{\sigma}_{\gamma\gamma\to m\bar{m}}(\hat{s}) = \frac{\pi \beta^5}{4\alpha^2\hat{s}}\left[\frac{3-\beta^4}{2\beta}\ln\frac{1+\beta}{1-\beta}-(2-\beta^2)\right]
\end{equation}
and for Drell Yan\cite{dougall}
\begin{equation}
    \hat{\sigma}_{q\bar{q}\to m\bar{m}}(\hat{s}) = \frac{\pi\eta_q^2\beta^3}{12\hat{s}}\left(2-\frac{2}{3}\beta^2\right).
\end{equation}
Once including the magnetic term, the cross sections can be recalculated\cite{baines} for photon fusion

\begin{equation}
    \label{ggkpaircrossec}
         \begin{split}
        \hat{\sigma}^\kappa_{\gamma\gamma\rightarrow m\bar{m}}(\hat{s},\kappa) &=\frac{\pi\alpha^2_m(\beta)}{3\hat{s}}\left\{ \ln\left(\frac{1-\beta}{1+\beta}\right)\left[\beta^2\kappa^2\hat{s}(3\beta^2\kappa^2\hat{s}-6\kappa^2\hat{s}+6)  +6\beta^4\right.\right. \\&\left.-(36\beta^2 -72\beta)\kappa\sqrt{(1-\beta^2)\hat{s}}        -9\kappa^4\hat{s}^2 -60\kappa^2\hat{s} -18
        \right]\\
        &\left.-\beta\kappa^2\hat{s}(7\beta^2\kappa^2\hat{s}^2+15\kappa^2\hat{s}+132)+ 12\beta^3 -24\beta-36\kappa\sqrt{(1-\beta^2)\hat{s}} \right\}
        \end{split}
\end{equation}

and Drell Yan
\begin{equation}
\label{dykpaircrossec}
    \hat{\sigma}^\kappa_{q\Bar{q}\to m\Bar{m}}(\hat{s},\kappa) =\frac{\pi\eta^2\beta^3}{18\hat{s}} \left[3-\beta^2-(2\beta^2-3)\kappa^2\hat{s}+6\kappa\sqrt{\hat{s}-\beta^2\hat{s}}\right].
\end{equation}

In the next section the formalism for the monopole-antimonopole bound state will be presented.

\subsection{Monopolium production}

The monopolium is a theorized bound state between a monopole and an antimonopole, first proposed in \cite{hill} as a possible relic of magnetic monopoles produced in the early universe. Due to the large magnetic coupling, a monopole-antimonopole pair would probably annihilate into a pair of photons\cite{garcia3}\cite{vento} or form a monopolium, which can have a lower mass and more stability. The monopolium can assume two spin values, $0$ and $1$, but in this work the spin $0$ monopolium is choosen to be studied in order to consider the lowest energy case.

The binding energy is written in terms of the monopole and  monopolium masses $m$ and $M$, respectively, $M = 2m + E_{binding}$, and to write the cross section for the photon fusion production the knowledge of the interaction potential in the pair is required. Considering the large coupling between the monopole and its antiparticle, it is possible to argue that both have some spatial extension\cite{schiff}, meaning that the interaction is non-singular when their separation ($r$) goes to zero. This can be described with the potential\cite{garcia2}
\begin{equation}\label{Mpot}
    V(r) = -g^2\left(\frac{1-e^{-\mu r}}{r} \right),
\end{equation}
where $\mu = 2m/g^2$ is a cut-off parameter that describes the interaction when $r\rightarrow 0$. The wave function can then be obtained by solving the Schrödinger equation with the interaction potential, and the cross section takes the form as \cite{garcia}
\begin{equation}\label{Ms}
     \hat{\sigma}_{\gamma\gamma\to M}(\hat{s}) =
     \frac{2\sqrt{2}[R(R-1)]^{3/2}}{\alpha^2\epsilon^6M^2}
     \frac{\bar{\Gamma}_M(\epsilon^2-1)^2}{(\epsilon^2-1)^2+\bar{\Gamma}_M^2},
\end{equation}
where $R=2m/M$, $\epsilon = \sqrt{\hat{s}}/M$ and $\bar{\Gamma}_M = \Gamma_M/M$ is the normalized decay rate of the monopolium. This cross section again can be used for processes that involve photon fluxes, like proton and nucleus collisions. However, it is not yet understood how the monopolium can be directly detected, and some models for the decay into two photons have been studied\cite{garcia3}\cite{vento} and give similar cross sections.

The collision of two ultrarelativistic protons can generate several processes involving photons, quarks and gluons and has a cleaner sign than a collision between a proton and a nucleus or two nucleus. Proton beams at LHC are also more energetic than the nucleus ones and can produce heavy particles up to a few TeV, which is the expected mass range of the monopole. A study of monopole production by photon fusion in nucleus collisions can be seen in \cite{sauter}, and as noted there, despite the enhancement of $Z^2$ due to the nucleus, the cross sections are very small ($\sim 10^{-10}$ fb) for monopole masses in the range $400$ to $1000$ GeV.

The usual formalism for proton-proton collisions can be done as presented in \cite{drees}, where the cross sections are written in a factorized form containing the cross section of the subprocess and the necessary distribution functions and photon fluxes.  The total cross section for a photon fusion process is written as the sum of the elastic, semielastic and inelastic contribution, while in Drell Yan the cross section is given by only one term, summed over all quark flavors. For the elastic photon flux of the proton it was used the expression given in \cite{drees2}, and the equivalent photon spectrum of quarks for the non-elastic terms and Drell Yan is given in \cite{drees}. For the proton structure function, it was used the Cteq6-1L parametrization \cite{cteq} with scale $Q^2 = \hat{s}/4$, to compare with the results in previous works\cite{dougall}\cite{garcia3}\cite{garcia}.

\subsection{Experimental search for monopoles}\label{subsec Experimental Search}

The current LHC experiments dedicated to the search of magnetic monopoles are the MoEDAL\cite{MoEDAL} experiment and the search for HIPs (highly ionizing particles) in ATLAS\cite{atlas2020}. The last results from ATLAS\cite{atlas2020} give a lower limit of $m \leq 2370$ GeV for a spin $1/2$ monopole with one or two units of magnetic charge ($n=1$ or $2$ in the quantization condition) produced via Drell-Yan, considering that the monopoles couple to photon with $\alpha_m\sim g^2$. The MoEDAL experiment evaluated the limits considering both Drell Yan and photon fusion production, and obtained a lower mass limit of $m \leq 2420$ GeV\cite{moedal2019} for the same type of coupling, and a limit of $m\leq 1760$ TeV using the velocity dependent coupling. So far there is no prediction regarding the magnetic moment dependent coupling. Taking these high limits into account, it is very probable that the search for magnetic monopoles will continue in future accelerators with higher collision energies and luminosities. For this reason the simulations are extended to the HE-LHC\cite{helhc} and FCC\cite{fcc} colliders, expected to begin their operations after 2035, and to the high luminosity LHC (HL-LHC)\cite{hllhc}, which will begin its operations in 2027 with the same energies of the LHC and a higher luminosity. 

In Table \ref{tabaccelerators} the main parameters of the pp colliders considered in the calculations are presented. The luminosity per year refers to the total luminosity of allocated physics time in a year, and corresponds to $160$ days for all accelerators. 

\begin{table}
\caption{Main parameters\cite{helhc}\cite{peakluminosity} of the LHC, HL-LHC, HE-LHC and FCC colliders. The beam energy is given in TeV, the peak luminosity in  \small{$10^{-5}$ fb$^{-1}/$s}, and the luminosity per year in $fb^{-1}$.}
\label{tabaccelerators}       
\begin{tabular}{lllll}
\hline\noalign{\smallskip}
\textrm{Parameters}&
\textrm{LHC}&
\textrm{HL-LHC}&
\textrm{HE-LHC}&
\textrm{FCC}  \\
\noalign{\smallskip}\hline\noalign{\smallskip}
 Beam Energy & 14 & 14 & 27 & 100 \\ 
Peak Luminosity & 1 & 5 & 16 & 5-30 \\  
Luminosity/year & 55 & 350 & 500 & 250-1000 \\
\noalign{\smallskip}\hline
\end{tabular}
\end{table}

\section{Results and Comments}\label{results}
\subsection{Monopole Production}\label{subsec results-monopole-production}
\begin{figure}
\centering
\resizebox{0.6\textwidth}{!}{%
  \includegraphics{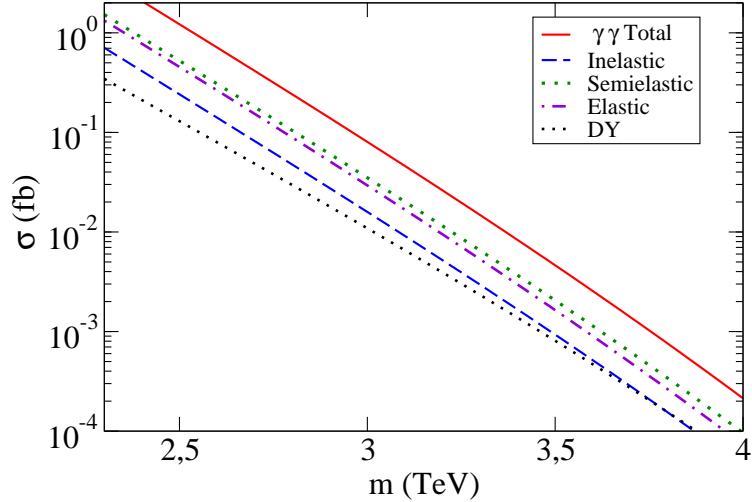}}
\caption{Monopole pair production via photon fusion and Drell Yan in pp collisions with $\hat{\kappa}=0$ and $\sqrt{s}=14$ TeV.}
\label{figmm}      
\end{figure}

\begin{figure}
\centering
\resizebox{0.6\textwidth}{!}{%
  \includegraphics{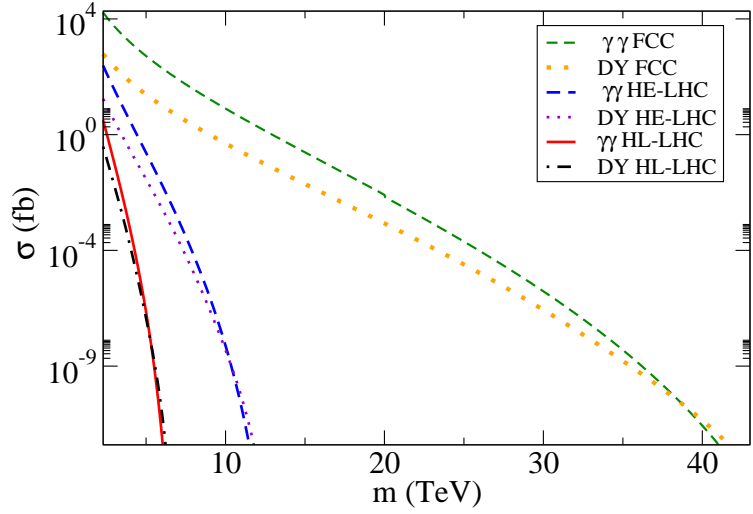}}
\caption{Monopole pair production via photon fusion and Drell Yan in pp collisions with $\hat{\kappa}=0$ at future accelerators.}
\label{figk0mmall}    
\end{figure}

The cross sections for the monopole-antimonopole pair production by photon fusion and Drell Yan ($\hat{\kappa} = 0$), at center of mass energy $\sqrt{s}=14$ TeV are presented in Fig. \ref{figmm}. The results corroborate with those in \cite{dougall}, \cite{garcia3} and \cite{baines}, meaning that the two-photon process has higher cross sections and it is a good candidate for simulations in LHC and new accelerators. For $m> 5500$ GeV the Drell Yan cross section overcomes the photon fusion, and this phenomenon also occurs in the HE-LHC and FCC calculations for $m> 10$ TeV and $m> 40$ TeV, respectively, as shown in Fig. \ref{figk0mmall}. However, considering the luminosities in Table \ref{tabaccelerators}, the cross sections for these masses are not relevant in any of the accelerators.

In Table \ref{tablimits}, the expected number of events per year for monopole production considering both Drell Yan and photon fusion are displayed. It can be seen that even for the LHC successor, the HL-LHC, monopoles with $m > 3$ Tev would probably not produce enough data to be confirmed. With the HE-LHC and FCC, monopoles with masses $m < 5$ TeV and $m < 18$ TeV, respectively, would have a higher probability to be detected.

\begin{table}
\caption{Number of events of monopole production (Drell Yan $+$ photon fusion) per year, for different monopole masses (in TeV).}
\label{tablimits}
\begin{tabular}{lllll}
\hline\noalign{\smallskip}
\textrm{Mass}&
\textrm{LHC}&
\textrm{HL-LHC}&
\textrm{HE-LHC}&
\textrm{FCC} \\
\noalign{\smallskip}\hline\noalign{\smallskip}
3 & $<10$ & $<40$ & $<3\cdot 10^4$ & $<2\cdot 10^7$ \\ 
5 & $< 2\cdot 10^{-5}$ & $<8\cdot 10^{-5}$ & $<150$ & $<2\cdot 10^6$ \\ 9 & 0 & 0 & $<4\cdot 10^{-4}$ & $<3\cdot 10^4$ \\
20 & 0 & 0 & 0 & $<10$ \\  
30 & 0 & 0 & 0 & $<2\cdot 10^{-3}$ \\
\noalign{\smallskip}\hline
\end{tabular}
\end{table}

\begin{figure}
\centering
\resizebox{0.6\textwidth}{!}{%
  \includegraphics{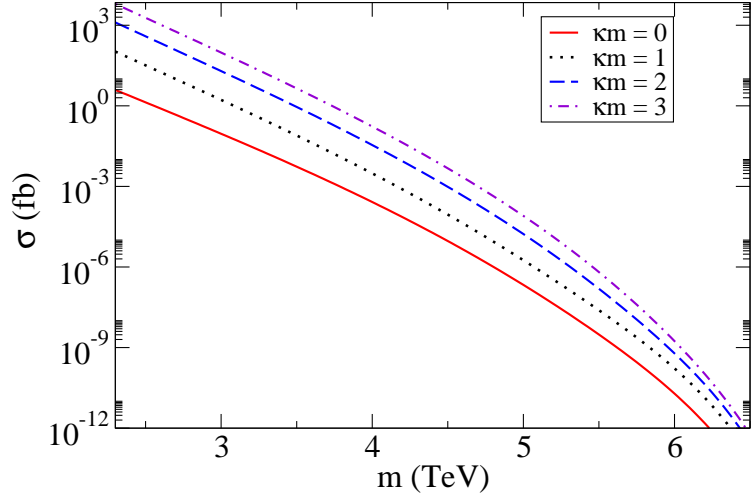}}
\caption{Monopole pair production considering both photon fusion and Drell Yan production in pp collisions for different values of $\hat{\kappa}$ and $\sqrt{s}=14$ TeV.}
\label{figkmm}    
\end{figure}

The total cross sections for pair production considering both photon fusion and Drell Yan are compared in Fig. \ref{figkmm} for different values of the magnetic moment parameter $\hat{\kappa}$, again for $\sqrt{s} = 14$ TeV. The results in \cite{baines} point out that the cross sections for the subprocesses and the total cross sections increase with the parameter $\kappa$, and the same behavior is achieved for the entire mass range. The cross section for $\hat{\kappa} =3$ is up to $10^2$ times higher than the one for $\hat{\kappa}=0$ only in photon fusion and around $10$ times higher in Drell Yan. It can then be concluded that the magnetic moment dependence, besides providing more applicability to the perturbation methods, can also increase the monopole detection chances.

\subsection{Monopolium production}\label{subsec results-monopolium-production}
\begin{figure}
\centering
\resizebox{0.6\textwidth}{!}{%
  \includegraphics{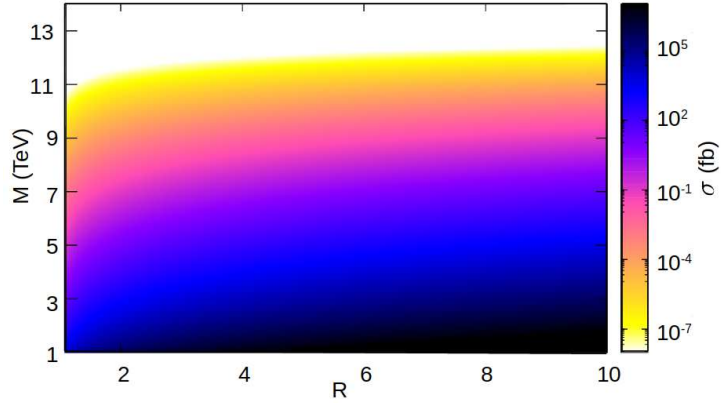}}
\caption{Monopolium production for different values of $M$ and $R = 2m/M$ in pp collisions with $\sqrt{s}=14$ TeV.}
\label{figM} 
\end{figure}

\begin{figure}
\centering
\resizebox{0.6\textwidth}{!}{%
  \includegraphics{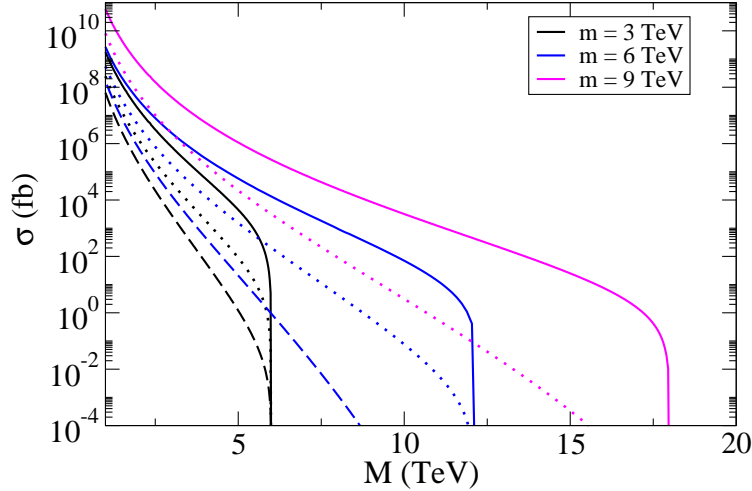}}
\caption{Monopolium production for fixed monopole masses in pp collisions at different accelerators (FCC: straight line, HE-LHC: dotted line, HL-LHC: dashed line).}
\label{figmall}
\end{figure}

The results for Monopolium production are in Fig. \ref{figM} and Fig. \ref{figmall}, and it can be seen that the cross section decreases with a lower rate when the monopolium mass is raised, compared to the monopole pair production. The production is also increased for higher values of monopole mass, supporting the results in \cite{garcia3}, \cite{garcia} and \cite{sauter}.

Considering a minimum of 1 event per year, the limits of detection in the LHC for monopolium production  are $M< 5$ TeV, for a fixed monopole mass of $m=3$ TeV. For the HE-LHC and FCC energies and luminosities this limit is close to a maximum possible mass, $M=6$ TeV. An analysis of the possible decay channels of the monopolium would improve the estimation of production and detection in a next step.

\section{Conclusions}\label{conclusions}

Based on these results, the production of magnetic  monopoles under the processes discussed here, makes it reasonable to expect that their experimental evidence can take a while. The lack of a well built perturbative theory to study the magnetic monopole in QED is still one of the greatest difficulties in obtaining predictions for their interactions. Our results confirm that the production by photon fusion is more relevant in the cross section range that allows detection in the LHC and future accelerators. Although also preliminary, the study of the magnetic moment term could lead to more applicability and new results to increase the chances of detection of such particles. The predictions for the mass range here obtained indicate that the production of monopoles in this model are at least one order of magnitude higher in comparison to the velocity dependent model.

The focus of this work is to restrict the limits of monopole production in view of the accessible energies of the current and next generation accelerators. Hence, it was opted to let GUT and other monopoles beyond the Standard model, such as those in \cite{twohiggs}, aside in the presented predictions, although they may be taken into consideration for future works. If the lower bounds in future results given experiments in accelerators continue to grow, this may indicate that one has to look for other possible monopole sources. 

\section*{Acknowledgments}\label{acknowledgments}
The authors would like to thank the Brazilian
funding agency Conselho Nacional de Desenvolvimento Científico e Tecnológico (CNPq) that partially financed this work.

\end{document}